\documentclass[preprint,showpacs,preprintnumbers,amsmath,amssymb]{revtex4}

\usepackage{graphicx}
\usepackage{dcolumn}
\usepackage{bm}
\usepackage{pstcol}

\newpsobject{showgrid}{psgrid}{subgriddiv=1,griddots=10,gridlabels=6pt}

\begin{document}


\title{Molecular Dynamics Simulation of Sympathetic Crystallization of Molecular Ions}

\author{Stephan Schiller}
\author{Claus L\"ammerzahl}%
\affiliation{%
Institute for Experimental Physics, Heinrich--Heine--University D\"usseldoerf, 40225 D\"usseldorf, Germany}%


\date{\today}

\begin{abstract}
It is shown that the translational degrees of freedom of a large variety of molecules, from light diatomic to heavy organic ones, can be cooled sympathetically and brought to rest (crystallized) in a linear Paul trap. The method relies on endowing the molecules with an appropriate positive charge, storage in a linear radiofrequency trap, and sympathetic cooling. Two well--known atomic coolant species, ${}^9{\hbox{Be}}^+$ and ${}^{137}{\hbox{Ba}}^+$, are sufficient for cooling the molecular mass range from 2 to 20,000 amu. 
The large molecular charge required for simultaneous trapping of heavy molecules and of the coolant ions can easily be
produced using electrospray ionization. Crystallized molecular ions offer vast opportunities for novel studies.
\end{abstract}

\pacs{42.50.Vk, 32.80.Pj, 33.80.Ps, 39.25.+k}
\maketitle

After the enormous success achieved in the field of cold atom manipulation,
significant efforts are under way to develop similar methods for molecules.
Samples of trapped ultracold molecules could be used for performing studies of
molecular structure, chemical reactions, quantum optics and molecular
Bose--Einstein condensates.

While methods for trapping of molecules, such as magnetic traps \cite{Weinstein98}, electrostatic traps \cite{Bethlem}, radiofrequency (Paul) or electromagnetic (Penning) traps \cite{Gosh95} or dipole traps \cite{Takeoshi} have been demonstrated and are in part already well developed, translational cooling of molecules is a field still under intense development. Direct laser cooling is not applicable due to lack of closed transitions. 
A technique demonstrated early on for cooling of neutral and charged molecules is by means of cryogenic buffer gas such as ${}^4$He, see e.g. \cite{buffergas1,buffergas2,BarlowDunnSchauer84}. Its extension to ${}^3\hbox{He}$ \cite{Weinstein98} allowed reaching temperatures below 1\,K. Two recently developed methods are the production of ultracold diatomic neutral molecules by photoassociation of ultra-cold atoms \cite{Pillet}, and the deceleration of polar molecules by time-dependent electric fields \cite{Meijer_dec}.

Another powerful method is sympathetic cooling  of ''sample'' particles of one species by an ensemble of directly cooled (often by laser cooling) particles of another species via their mutual interaction. This technique, first demonstrated for ions in Penning traps \cite{NIST1,NIST2}, is being applied to an increasingly wide variety of neutral and charged particles (atoms, molecules, elementary particles) in various trap types for applications ranging from mass spectrometry to quantum computing. 

In both Penning and radiofrequency traps, first studies showed that molecular ions could be sympathetically cooled
(sc) by laser-cooled (lc) atomic ions to temperatures in the range of several K \cite{earlySC_mol1,earlySC_mol2}, and
mass ratios down to $m_{\rm sc}/m_{\rm lc} = 2/3$ were achieved \cite{BabaWaki01}. An important aspect of ion traps
is that for sufficiently strong cooling the formation of an ordered structure (Coulomb crystal) results
\cite{Wakietal92,Raizenetal92}. Recently it has been shown that sympathethic crystallization of molecular ions is
possible in a linear rf trap where the molecules are stably incorporated into the atomic Coulomb crystal
\cite{MoelhaveDrewsen00,Hornekaer00}. The molecular ions included MgH$^+$, MgD$^+$ (cooled by ${\hbox{Mg}}^+$) and
${}^{16}{\hbox{O}}_2^+$ (cooled by ${}^{40}{\hbox{Ca}}^+$ or ${}^{24}{\hbox{Mg}}^+$), i.e. mass ratios down to 0.6. In
sympathetic crystallization of atomic ions, a mass ratio range of 0.8 - 1.8 has been achieved \cite{Kai,Drewsen03}.

The range of molecular mass that can be sympathetically cooled to temperatures of the order of 10\,K, where the ion ensemble is still in a gas state, has very recently been debated in theoretical work \cite{BabaWaki02,BabaWaki02a,Harmonetal02}. A molecular mass range from 8 to 192 amu (mass ratios 1/3 - 8) was found using molecular dynamics simulations to be accessible using ${}^{24}{\hbox{Mg}}^+$ as coolant ion.

In the present simulations we study a much larger mass range and the regime of much 
lower temperatures ($\ll 1\,$K) and study whether sympathetic crystallization can be reached. We find that essentially
all molecular masses can be sympathetically crystallized by one of two commonly used species of laser--coolable ions,
${}^9{\hbox{Be}}^+$ and ${}^{137}{\hbox{Ba}}^+$. The only requirement for this general method is an appropriate charge
state for the molecules. Single positive charges are sufficient for small molecules (mass 2 -- 2000 amu). For heavier
molecules  higher positive charge states are required in order to allow for reliable simultaneous trapping. These are
easily produced using electrospray ionization \cite{Fenn}. The ability to store molecules in an almost motionless state in a
collision-free ultra-high vacuum environment for essentially unlimited
time is expected to open up vast opportunities for high precision
spectroscopy and the study of slow molecular processes.

In order to model sympathetic cooling in rf traps, it is crucial to take collisions into account precisely. Noninteracting ions perform an oscillation at the frequency of the applied rf field, but their cycle-averaged energy is constant if ion--ion interactions can be neglected.  When interactions, i.e. collisions, between the ions are taken into account, energy gain from the rf field (rf--heating) and transfer of energy from one species to another, i.e. sympathetic heating and cooling, occurs. Approximate models have been proposed to describe these mechanisms \cite{BabaWaki02,BabaWaki02a}. However, it is highly desirable to perform calculations that are essentially free of approximations. We therefore choose a molecular dynamics 
(MD) approach \cite{Prestageetal91,Schifferetal00}. 

The simulations are based on solving Newton's equations of motion for the laser--cooled (lc) and sympathetically cooled ions (sc)
\begin{equation}
m_i \ddot{\bf x}_i = Q_i {\bf E}_{\rm trap}({\bf x}_i,t)+{\bf F}_{{\rm C},i}(\{{\bf x}_j\})+{\bf
F}_{\rm L}(\dot{\bf x}_i,t) \, ,
\end{equation}
where $i=1,\ldots, N_{\rm lc}+N_{\rm sc}$ ($N_{\rm lc}$ and $N_{\rm sc}$ are the numbers of 
laser--cooled and sympathetically cooled particles, respectively). Positions, charges and masses are ${\bf x}_i$,
$Q_i$, and $m_i$, the Coulomb force ${\bf F}_{{\rm C},i}=(Q_i/4\pi\epsilon_0)\nabla_i\sum_{j}Q_j/r_{ij}$, where
$r_{ij}$ is the distance between particles $i$ and $j$. Here ${\bf E}_{\rm trap} = \mbox{\boldmath$\nabla$}((x^2-y^2)
\cos(\Omega t)V_{\rm rf}/2r_0^2+(z^2-x^2/2-y^2/2)U_{\rm dc}/d^2)$ is the electric field in a linear ion trap
\cite{Gosh95}, where $z$ is along the trap axis. The radial and axial dimensions of the trap are $r_0$ and $d$. A
necessary condition for stable trapping of noninteracting ions is a $q$--parameter, $q_i = 2 (Q_i/m_i) V_{\rm
rf}/(\Omega^2 r_0^2)$ less than 0.9. However, it is well-known experimentally and theoretically that operation at
significantly smaller $q$ is favourable since rf--heating is less pronounced \cite{Harmonetal02,Prestageetal91}. On the
other hand, a lower limit is given by experimental considerations, $q_{\rm min} \simeq 0.05$. We therefore choose
$q$--parameters in the range 0.05 -- 0.4 for both atomic and molecular ions. As a consequence, the simultaneous storage
of heavy molecules ($m_{\rm sc}
\gg m_{\rm lc}$) requires a molecular charge exceeding unity. 

Laser cooling of the lc particles is described by the force ${\bf F}_{\rm L}$.
In actual experiments, its strength is such that cooling may require minutes. Such durations correspond to
$10^8$ or more rf periods, and cannot be simulated in high--precision MD when
particle numbers are large. In order to compromise between a reasonable
computing time and realistic laser cooling strength we have used 
stronger forces ${\bf F}_{\rm L}$ to speed up cooling. Most simulations were
performed with a simple linear viscous damping ${\bf F}_{\rm L}=-\beta \dot{\bf
x}$ with friction coefficients in the range $\beta = (1.2 - 8)
\cdot10^{-22}$kg/s. These are well below the maximum value (at optimum laser
detuning from the cooling transition), $\beta_{\rm
max}\simeq\pi^2\hbar/\lambda^2\simeq$ $4\cdot10^{-21}$kg/s for transitions at
optical wavelengths $\lambda$. For some simulations, we have used more realistic forces, see below. 

The diffusion of the lc ion momenta in momentum space due to recoils upon spontaneous emission gives rise to the Doppler cooling limit. This diffusion is included in the simulations. 

The equations of motion are solved using a high--order Runge--Kutta method with adaptive step size. 
Initial conditions were in part chosen so as to give initial temperatures below room
temperature, again in order to reduce the computational effort.

In order to characterize the state of the plasma, from the trajectories of the particles we calculate, 
for each species, the average kinetic energy per particle, $E_{\rm kin}^{\rm tot} = (2 N_\kappa)^{-1}\sum_\kappa m_i
\langle {\bf v}_i^2(t) \rangle$, where $\langle\cdot\rangle$ is the time--average over one period of the rf field, the
time--averaged secular energy per particle $E_{\rm kin}^{\rm sec} = (2 N_\kappa)^{-1}\sum_\kappa {m_i|\langle {\bf
v}_i(t)\rangle|^2}$, and the average interaction energy per particle (at the end of a rf period). Here $\kappa$ means
lc or sc, respectively, and $\sum_\kappa$ denotes  summation over the corresponding species. The secular energy, where
the micromotion oscillation is averaged out,  can be taken as an indication of the temperature of the sample, since it
arises from the "disordered" motion of the interacting ions in the time--averaged trapping potential. In contrast, the
micromotion contribution to the total energy arises from a regular motion. We note that in simulations where the rf
potential is replaced by the pseudopotential, i.e. the time--averaged trap potential $V_{\rm pseudo}(\rho)=  \rho^2 Q^2
V_{\rm rf}^2/(4 m \Omega^2 r_0^2)$ experienced by the particles in radial direction, the cooling is moderately faster.
This implies that rf heating is nonnegligible even at the relatively low $q$--values chosen here.

In the following we describe three mass regimes of sympathetic
crystallization.

\paragraph*{Cooling of molecular hydrogen isotopomers}

One challenging goal in ultracold molecule studies 
will be the precision spectroscopy of the simplest (i.e. one-electron)
molecules, the hydrogen ions ${\hbox{H}}_2^+$, ${\hbox{HD}}^+$, ${\hbox{D}}_2^+$. The experimental accuracy
can potentially be improved by several orders, surpassing by far the
current theoretical precision \cite{27neu}. It will then become possible to
test and challenge calculation methods, especially of the relativistic and QED
contributions. Moreover, since
the vibrational energies depend explicitly on the electron--to--proton mass ratio $m_e/m_p$ \cite{27b},
their measurement might allow to
determine the value of this fundamental constant by 
spectroscopic means, providing an alternative and potentially more
accurate approach than mass measurements in Penning traps \cite{Beieretal02}.


A precise measurement will require cold trapped molecules in order to minimize Doppler broadening. The heteronuclear ${\hbox{HD}}^+$ is of particular interest since it has dipole--allowed vibrational transitions 
\cite{Wingetal76} that could be excited by infrared lasers such as optical parametric oscillators \cite{Kovalchuk01} or
diode lasers. 

A simulation of sympathetic cooling of 5 ${\hbox{HD}}^+$--ions by 20 laser-cooled ${}^9{\hbox{Be}}^+$--ions is shown 
in Fig.1. The laser quickly cools the  ${\hbox{Be}}^+$--ions to a  liquid state, characterized by comparable Coulomb
interaction energy and secular energy (plasma parameter $\Gamma \sim 2$). The atomic ion temperature remains constant
while the molecular ions are sympathetically cooled. Only once the molecular secular energy becomes comparable to the
atomic secular energy does the latter decrease further. The secular energy of the ${\hbox{Be}}^+$--ions finally reaches
the Doppler limit. The secular energy of the molecules reaches that level significantly more slowly, since the cooling
power of the atomic ions becomes smaller as they settle into the crystalline state. 

\begin{figure}[b]
\noindent\begin{pspicture}(0,0)(12,12)
\put(0,4.6){\includegraphics[width=12.5cm]{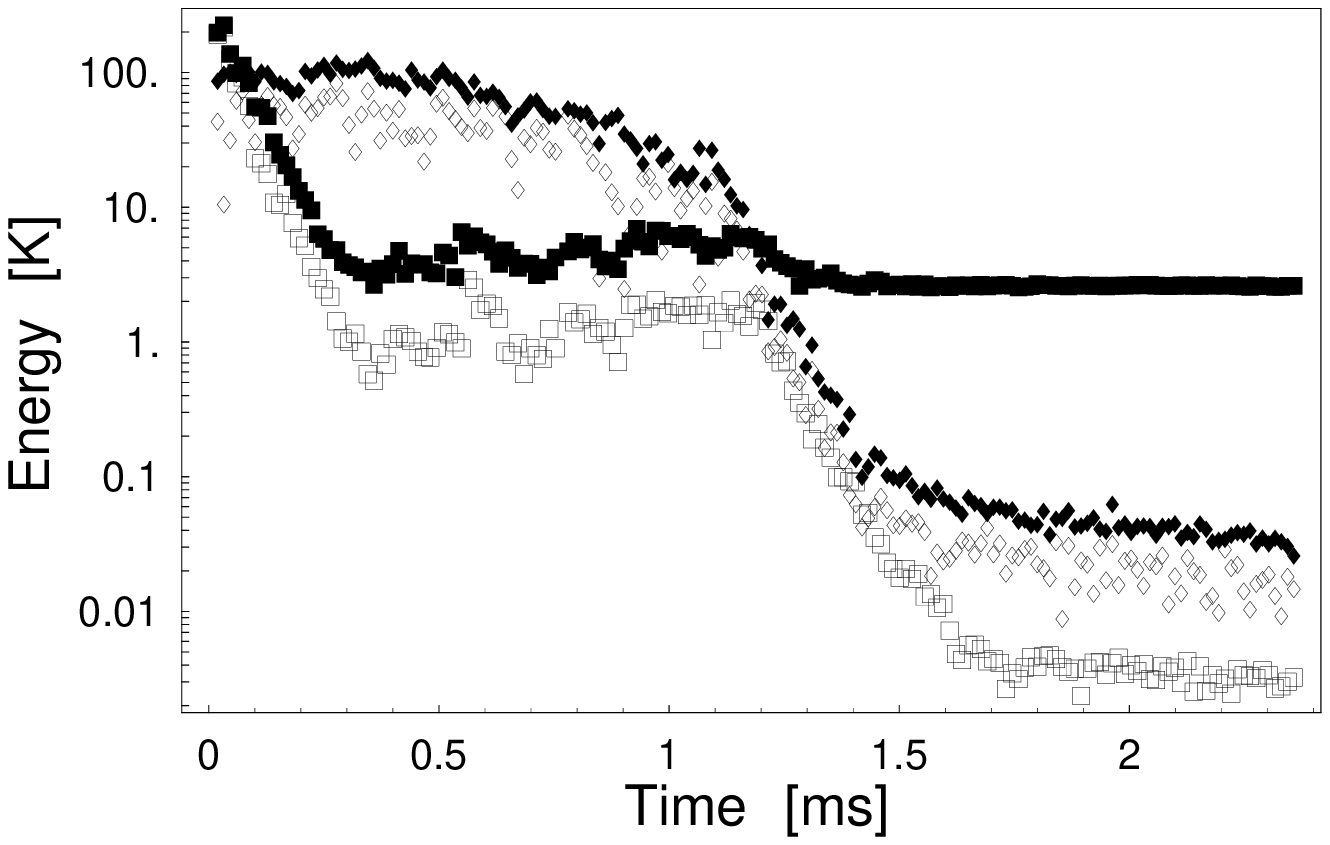}}
\rput(10,10.1){\large $E_{\rm kin}^{\rm tot}({}^{9}{\hbox{Be}}^+)$}
\rput(6.8,7){\large $E_{\rm kin}^{\rm sec}({}^{9}{\hbox{Be}}^+)$}
\rput(9,11){\large $E_{\rm kin}^{\rm tot}({\hbox{HD}}^+)$}
\rput(10.9,8.5){\large $E_{\rm kin}^{\rm sec}({\hbox{HD}}^+)$}
\psline{->}(7.8,11)(6.5,10.7)
\psline{->}(9.7,8.5)(9,7.3)
\put(1.4,0){\includegraphics[width=11cm,trim= 0 0 30 0,clip]{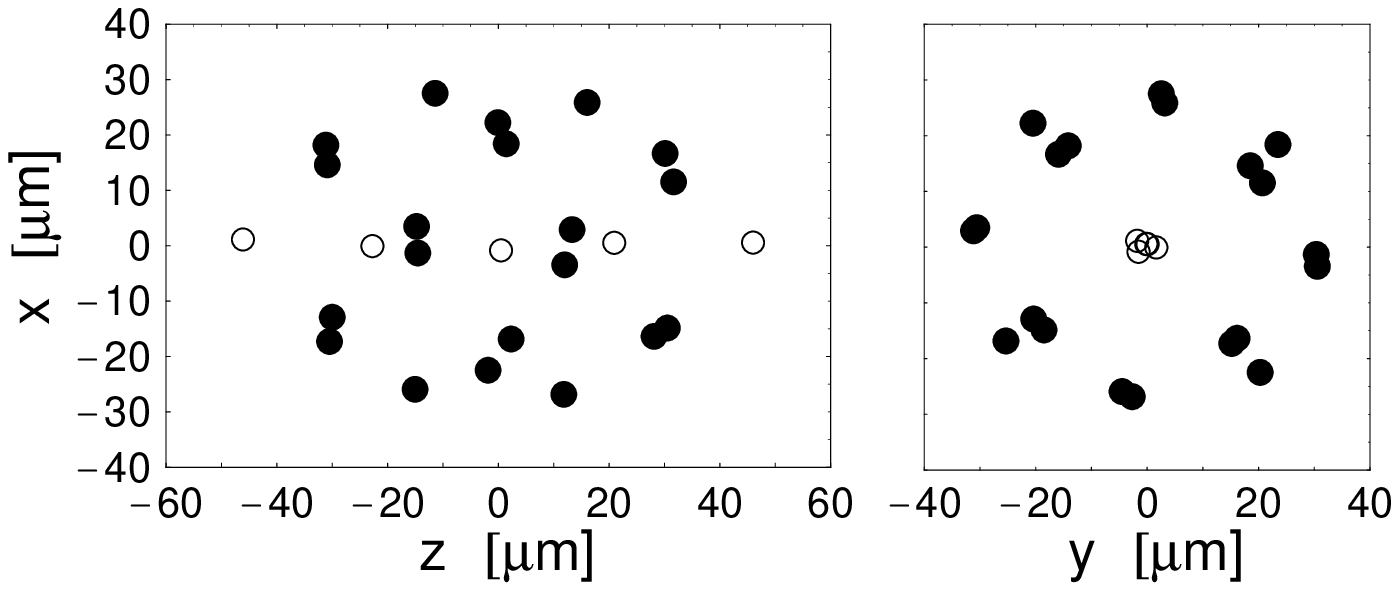}}
\put(0,12){\large a)}
\put(0,3.9){\large b)}
\end{pspicture}
\vspace*{-3ex}
\caption{ 
Simulation with $N_{\rm lc} = 20$ laser-cooled ${}^9{\hbox{Be}}^+$--ions and $N_{\rm sc}=5$ ${\hbox{HD}}^+$--molecules. {\bf a:} Energies per particle. 
{\bf b:} Spatial structure in the crystalline state (left: projection onto $x$--$z$--plane, right: $x$--$y$--projection); solid  circles: ${}^9{\hbox{Be}}^+$, circles: ${\hbox{HD}}^+$. 
The trap parameters are 
$q_{\rm lc} = 0.13$ 
$q_{\rm sc}=0.39$, 
rf frequency $\Omega/2\pi=8.5$\,MHz, 
$V_{\rm rf}/r_0^2=17.6\,\hbox{V}/{\hbox{mm}}^2$,
$U_{\rm dc}/d^2 =30\,\hbox{V}/{\hbox{cm}}^2$.
Viscous laser cooling with $\beta=2.4\times10^{-22}\hbox{kg/s}$ was used.
}
\label{fig1}
\vspace*{.0in}
\end{figure}

The spatial structure resulting from the cooling (Fig.~\ref{fig1}b) can be 
understood by considering the different pseudopotentials felt by the two particle species \cite{Hornekaeretal01}. It is
three times larger  for the lighter
${\hbox{HD}}^+$--ions as compared to the ${}^9{\hbox{Be}}^+$--ions. The total energy of the ensemble is minimized if the
${\hbox{HD}}^+$--molecules lie on-axis. The ${\hbox{Be}}^+$--ions form a shell structure around them. It is moderately
prolate, since the radial  pseudopotential and axial potential are similar in strength (${\hbox{Be}}^+$--oscillation
frequencies
$\omega_\rho/2\pi=340\,$kHz, $\omega_z/2\pi=285\,$kHz). The axial arrangement of the ${\hbox{HD}}^+$--ions is
favourable for spectroscopic investigations, since on--axis the micromotion is zero, with a corresponding
simplification of their transition spectrum. Since the ${\hbox{Be}}^+$--ions form a three--dimensional structure, their
micromotion energy remains relatively high compared to their secular energy (Fig.\ref{fig1}a). This is because the
off--axis locations of the ions imply corresponding micromotion velocities, proportional to the radial distances. 

Simulations with a more realistic (i.e. weaker) cooling force 
\footnote{The semiclassical cooling force with 
full dependence on particle velocity and detuning of the laser 
frequency from resonance \cite{Metcalf} was used, but
enhanced by a factor $\sim 8$. The detuning was repetitively scanned in a sawtooth manner from far red--detuned to
$-\gamma/2$ where $\gamma$ is the cooling transition linewidth. The laser beam direction was at 54.7\,$\deg$ with
respect to the trap symmetry axes.} were also performed. Qualitatively, the same behaviour resulted, however, as
expected, sympathetic crystallization was reached after a substantially longer time, about 10 times slower than in
Fig.\ref{fig1}a. 

A sample of molecular ions is not always sympathetically cooled in its entirety.
For example, in a simulation of 40 ${\hbox{Be}}^+$--ions and 10
${\hbox{HD}}^+$--ions, the final state of the system contained 6 on-axis
crystallized ${\hbox{HD}}^+$--ions embedded in a prolate
${\hbox{Be}}^+$--crystal, while 4 ${\hbox{HD}}^+$--ions remained hot. It is
clear that for small particle numbers the number of cooled particles depends on
details of the initial conditions. 

The above simulations for ${\hbox{HD}}^+$ are
of course also applicable to the astrophysically important ${\hbox{H}}_3^+$.
We have also performed simulations for other hydrogen ions. We have
found (partial) sympathetic crystallization for all masses, from 2 amu (${\hbox{H}}_2^+$) to 5 amu (${\hbox{DT}}^+$).

\paragraph*{Cooling of dye molecules}

Dye molecules are interesting model systems for studies of complex (i.e. polyatomic) ultracold molecule manipulation because they are well--characterized and can easily be excited optically. Various dye molecules have masses exceeding a few times the mass of heavy atoms. We consider here the case of cooling Rhodamine 6G, a molecule of mass 493 that can be transferred to the gas phase singly charged by means of electrospray ionization. The coolant ion is chosen to be ${}^{137}{\hbox{Ba}}^+$. In this case, the larger mass--to--charge ratio of the molecules leads to a shell structure with the atomic ions in the center and on--axis. The small fraction of the molecules which is well embedded in the atomic ensemble cools and crystallizes on the same timescale as the atomic ions. The remainder experiences a much weaker cooling due to the absence of a ''cageing'' effect. The timescale for cooling and crystallization of all molecules was found to be 10 times larger than for the atomic ions. 

\paragraph*{Cooling of large molecules}

\begin{figure}[t]
\begin{center}
\psset{unit=1cm}
\begin{pspicture}(0,0)(12,13)
\put(-0.2,5.5){\includegraphics[width=12.5cm]{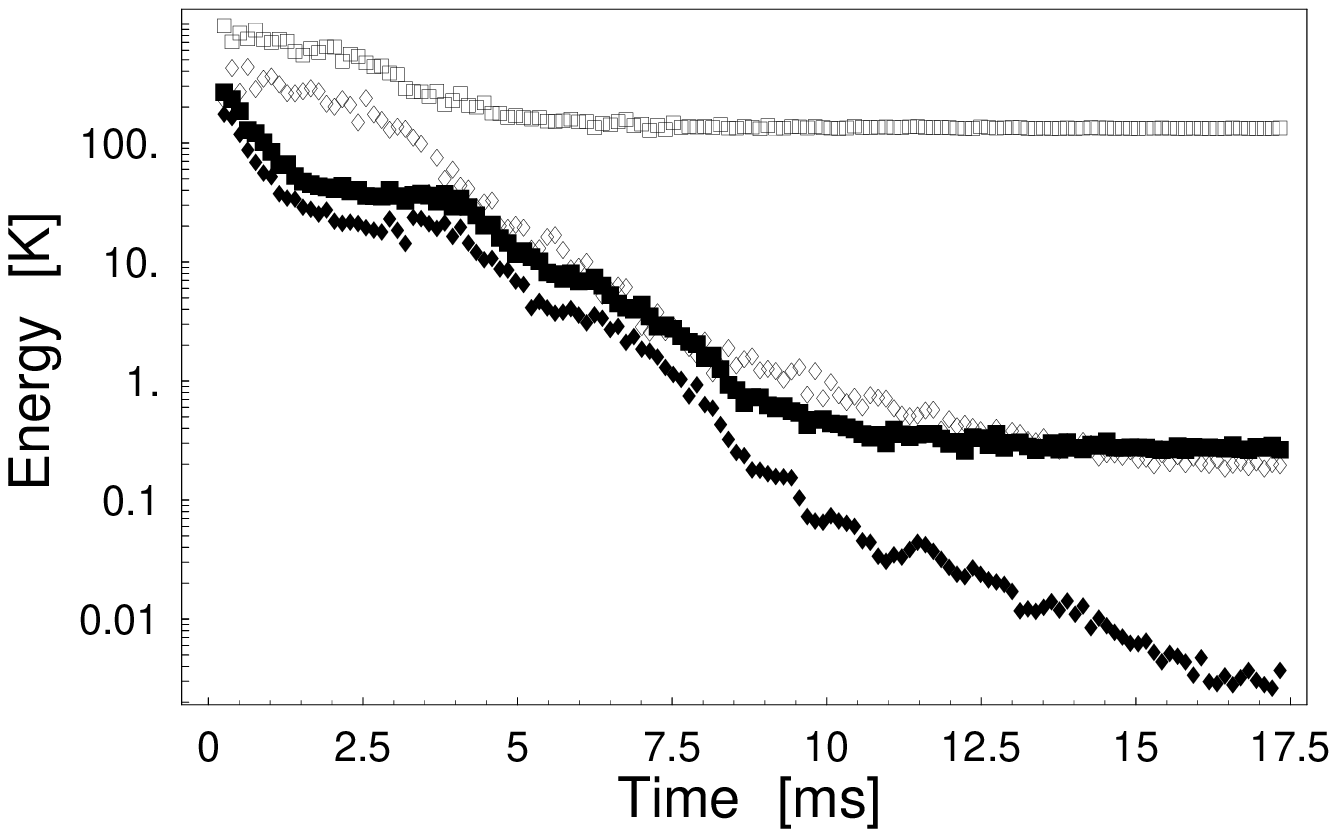}}
\rput(10.2,8.4){\large $E_{\rm kin}^{\rm tot}({}^{137}{\hbox{Ba}}^+)$}
\rput(7,7.5){\large $E_{\rm kin}^{\rm sec}({}^{137}{\hbox{Ba}}^+)$}
\rput(10,11.5){\large $E_{\rm kin}^{\rm tot}(\hbox{mol})$}
\rput(7.8,10.2){\large $E_{\rm kin}^{\rm sec}(\hbox{mol})$}
\psline{->}(8.8,8.4)(8.2,9)
\psline{->}(6,7.9)(6.8,8.7)
\put(0.02,3){\includegraphics[width=9.9cm]{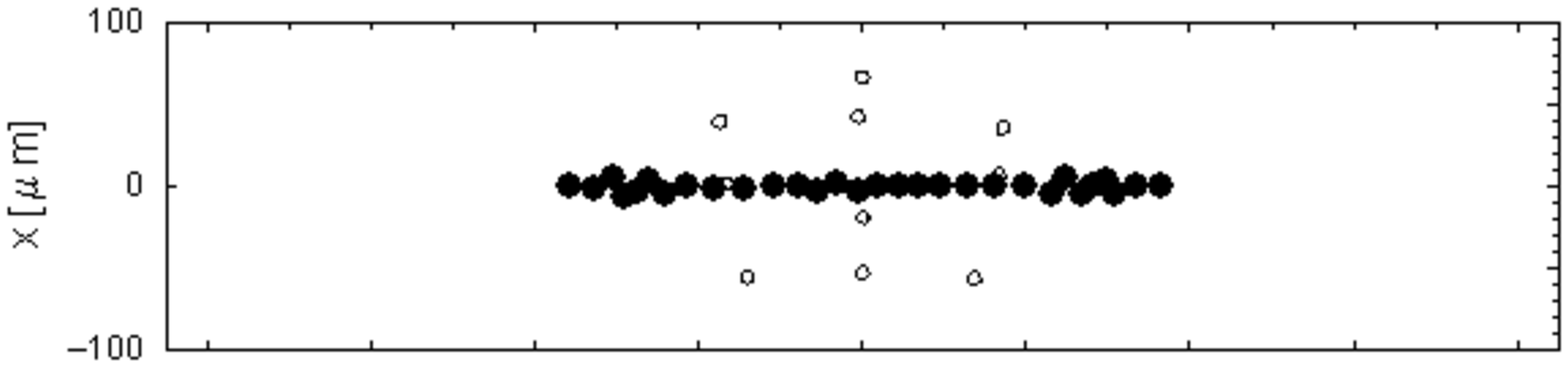}}
\put(10,3.03){\includegraphics[width=2.545cm]{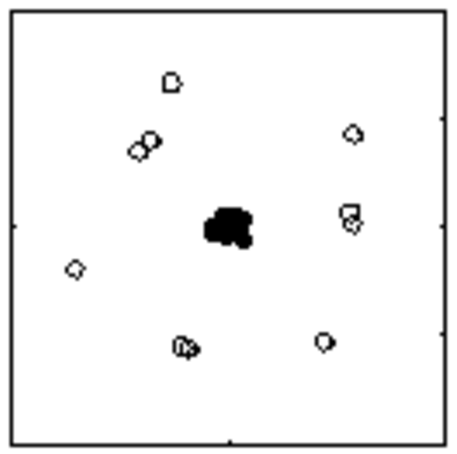}}
\put(0,0){\includegraphics[height=3cm]{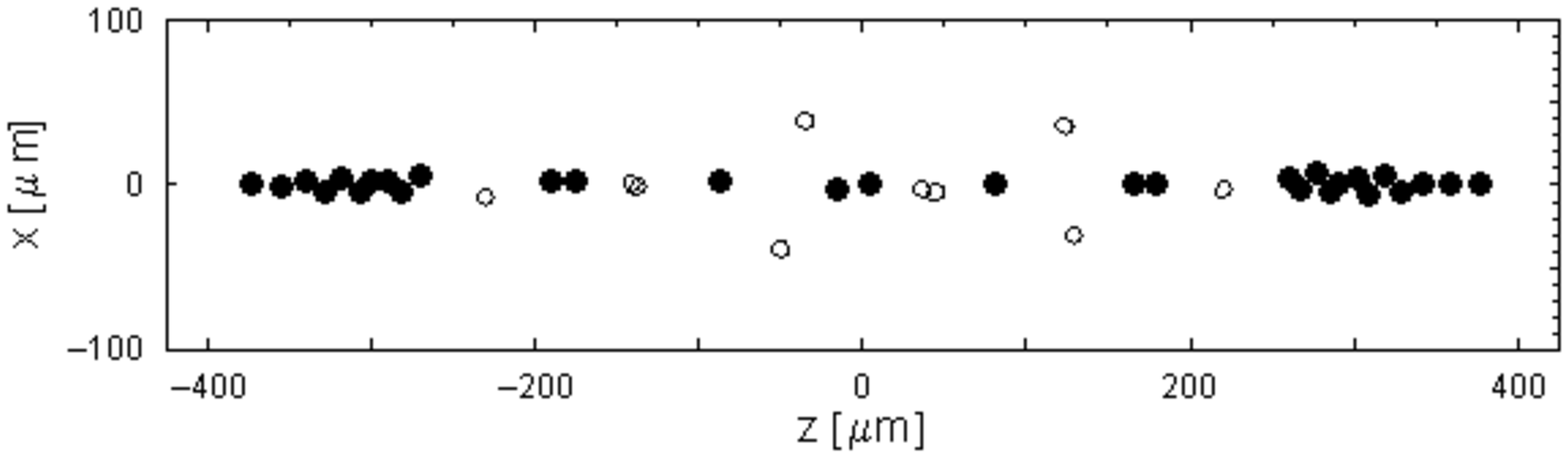}}
\put(10,0.05){\includegraphics[height=2.91cm]{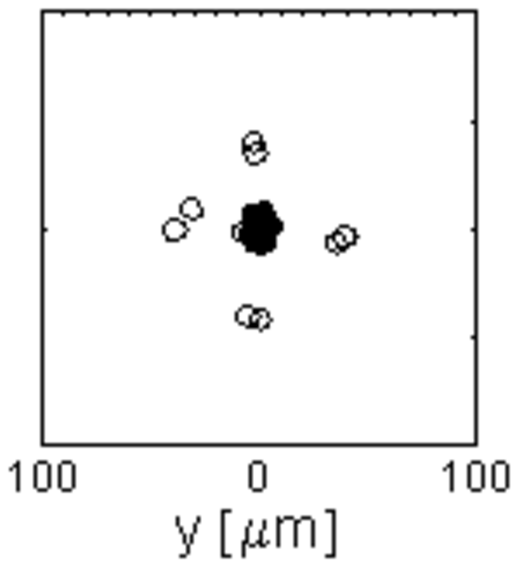}}
\pspolygon[linecolor=white,fillstyle=solid,fillcolor=white](0,0)(12.5,0)(12.5,0.7)(1,0.7)(1,5.3)(0,5.3)
\rput(01.2,0.45){\sffamily\footnotesize -400}
\rput(9.5,0.45){\sffamily\footnotesize 400}
\rput(5.4,0.2){\sffamily $z \;[\mu\hbox{m}]$}
\rput(10.3,0.45){\sffamily\footnotesize -100}
\rput(12.3,0.45){\sffamily\footnotesize 100}
\rput(11.25,0.2){\sffamily $y \;[\mu\hbox{m}]$}
\rput(0.55,0.9){\sffamily\footnotesize -100}
\rput(0.55,2.8){\sffamily\footnotesize 100}
\rput(0.55,3.2){\sffamily\footnotesize -100}
\rput(0.55,5.1){\sffamily\footnotesize 100}
\rput{90}(0.4,1.8){\sffamily $x \;[\mu\hbox{m}]$}
\rput{90}(0.4,4.1){\sffamily $x \;[\mu\hbox{m}]$}
\put(-0.4,13){\large a)} 
\put(-0.4,5.5){\large b)} 
\put(-0.4,2.8){\large c)} 
\end{pspicture}
\vspace*{-3mm}
\end{center} \caption{
Sympathetic cooling of 10 molecular ions of mass 20000 and charge $20\, e$ by 30 ${}^{137}{\hbox{Ba}}^+$--ions. (a) Energies.  
(b) Spatial structure.
(c) Spatial structure for a simulation where the axial potential is weaker. Parameters: $q_{\rm lc} = 0.33$
$q_{\rm sc}=0.045$, $\Omega/2\pi=1.6\;\hbox{MHz}$,
$V_{\rm rf}/2r_0^2=23.8\,\hbox{V}/{\hbox{mm}}^2$,
$U_{\rm dc}/d^2=12\,\hbox{V}/{\hbox{cm}}^2$ for (b), $3\,\hbox{V}/{\hbox{cm}}^2$ for (c).
Viscous laser cooling with $\beta=4.8\times10^{-22}\hbox{kg}/\hbox{s}$ was used.
}
\label{fig_large_molecules}
\vspace*{-4mm}
\end{figure}

Large molecules such as amino acids or proteins can be transferred into vacuum in high charge states by
electrospray ionization \cite{Fenn}. In this method, protons are attached to molecules that emerge from a solution
spray. Large molecules are produced with a distribution of charge states, with
typical specific charges in the range $m/Q=700 - 1500$. For example, the charge
states for a protein of mass 17,000 can range from 11 to 25 \cite{Cole97}. Such
specific charges are sufficiently near those of laser--coolable heavy atomic
ions (${}^{137}{\hbox{Ba}}^+$, ${}^{172}{\hbox{Yb}}^+$,
${}^{199}{\hbox{Hg}}^+$). Since the $q$--parameters $\sim Q_i/m_i$ are within an
order of magnitude, simultaneous trapping is possible. Moreover, the large charge
ensures a good coupling between molecular ensemble and atomic ensemble. Finally,
the pseudopotential of the molecular ions ($\sim Q_{\rm sc}^2/m_{\rm sc}$) can
be  comparable or even larger than that of the atoms, thanks to the large
molecular charge. A good spatial overlap of the subensembles is therefore ensured.
Fig.\ref{fig_large_molecules} shows
two simulations for molecular ions of mass 20,000 amu and charge 20\,$e$. As can be seen, the
molecular ions are strongly cooled as far as their secular energy is
concerned. However, a large micromotion energy remains in the crystallized state because of the nonaxial arrangement 
of the molecular ions and their large mass. This nonaxial arrangement occurs in spite of the fact that the
pseudopotential of the molecules is larger than that of the atoms. Indeed, the Coulomb interaction energy is an
important contribution that influences the overall arrangement of the particles. This can be seen by noting that the
Coulomb energy of a given spatial structure is not invariant under particle exchange lc$\leftrightarrow$sc, due to the
differing molecular and atomic charges. For a sufficiently large axial potential (Fig.\ref{fig_large_molecules}a,b) the
total energy is minimized by forcing the molecular ions off axis resulting in a shell structure. For smaller axial
potential (Fig.\ref{fig_large_molecules}c) the structure is string--like, but molecular ions that are adjacent bulge
out of the axis because of their strong repulsion. Where several molecular ions are crystallized in adjacent positions,
they form sections of a helix
\cite{HasseSchiffer90}. Here again we find that molecules that are well embedded in the atomic ion ensemble, i.e. those
that are individually located between atomic ions (e.g. the two sc ions in Fig.\ref{fig_large_molecules}c furthest from
the center in z-direction), cool as fast as the lc ions while the off-axis molecular ions cool much more slowly.

We have also performed simulations for molecular masses 2000, 5000, and 10,000 amu, with the same mass to charge 
ratio of 1000 amu/$e$. We find a similar behaviour of crystallization as in the examples above. As the charge becomes
smaller only molecular Coulomb structures of the shell--type occur, and due to the weaker lc--sc interaction, a
fraction of the molecules remain uncooled. 

The above examples show that sympathetic crystallization is possible for ion masses significantly 
larger or smaller than the coolant ion mass. Based on the present and previous \cite{Harmonetal02} results, it can be
stated that simply charged molecular ions of mass between that of ${}^9{\hbox{Be}}^+$ and ${}^{137}{\hbox{Ba}}^+$ can
also be sympathetically crystallized.

In conclusion, we have shown that a wide variety of charged molecules can be trapped and cooled to mK 
temperatures with masses from 2 to 20,000 amu, i.e. from diatomic molecules to polymers and proteins. The molecules are
incorporated into an ordered structure. The spatial arrangement within the Coulomb crystal depends on the masses and
charges of the coolant and the molecular ions. We have pointed out that sympathetic crystallization should be very
advantageous for precision spectroscopy of the various ions of molecular hydrogen. 

Since emphasis was placed on using a laser cooling force of realistic magnitude and on employing a
 high--precision numerical code, the simulations were performed for small particle numbers (up to 50). Experimentally, this regime is both accessible and suited for detailed
studies of molecular properties. In future, the
simulations can be extended to larger numbers, by developing faster but approximate algorithms or using more powerful
computers, and to other types of traps (multipole-rf traps \cite{buffergas2} and Penning traps). This should allow
detailed comparisons of dynamics and structure of sympathetic crystallization between theory and experiments currently
under way. 

\paragraph*{Acknowledgment}

We thank M. Drewsen, E. G\"okl\"u, M. Hochbruck and V. Grimm for stimulating discussions and D. Schumacher for providing computer support.

\end{document}